\documentclass[conference,a4paper]{IEEEtran}

\usepackage[utf8]{inputenc} 
\usepackage[T1]{fontenc}
\usepackage{url}              % provides \url{...}
\usepackage{cite}             % improves presentation of citations

\usepackage[cmex10]{amsmath}  % Use the [cmex10] option to ensure complicance
\usepackage{mleftright}       % fix to wrong spacing of \left-,
\mleftright                   % \middle- \right-commands 

\usepackage{graphicx}         % provides \includegraphics{...} to
\usepackage{booktabs}         % fixes poor spacing in tables and
\usepackage{xcolor}
\usepackage{amsfonts}
\usepackage{dsfont}
\usepackage{multirow}
\usepackage{tikzit_dep/tikzit}
\tikzstyle{OrangeBox}=[fill={rgb,255: red,255; green,192; blue,83}, draw=black, shape=rectangle, align=center]
\tikzstyle{BlueBox}=[fill={rgb,255: red,80; green,171; blue,255}, draw=black, shape=rectangle, align=center]
\tikzstyle{RoundedOrangeBox}=[fill={rgb,255: red,255; green,204; blue,114}, draw=black, shape=rounded rectangle, align=center]
\tikzstyle{RoundedBlueBox}=[fill={rgb,255: red,116; green,190; blue,255}, draw=black, shape=rounded rectangle, align=center]
\tikzstyle{WhiteBox}=[fill=white, draw=black, shape=rectangle, align=center]
\tikzstyle{Empty}=[fill=white, draw=none, shape=rectangle, inner sep=0cm, align=center]

\tikzstyle{thickLine}=[-, line width=1pt]
\tikzstyle{thickOrangeLine}=[-, fill=none, line width=1pt, draw={rgb,255: red,255; green,192; blue,83}]
\tikzstyle{thickBlueLine}=[-, draw={rgb,255: red,80; green,171; blue,255}, line width=1pt]
\tikzstyle{thickArrow}=[-, -latex, line width=1pt]

\newtheorem{thm}{Theorem}[section]
\newtheorem{lem}[thm]{Lemma}

\newtheorem{cor}{Corollary}
\newtheorem{defn}{Definition}[section]

\newcommand{\norm}[1]{\left\Vert #1 \right\Vert}
\newcommand{\twoline}[2]{\genfrac{}{}{0pt}{}{#1}{#2}}

\newcommand{\ket}[1]{| #1 \rangle}
\newcommand{\bra}[1]{\langle #1 |}

\DeclareMathOperator{\R}{\mathbb R}
\DeclareMathOperator{\tr}{\operatorname{tr}}

\DeclareMathOperator{\cH}{\mathcal H}
\DeclareMathOperator{\cS}{\mathcal S}
\DeclareMathOperator{\cB}{\mathcal B}

\newcommand{\identity}{\ensuremath{\mathds{1}}}

\begin{document}

\title{General Continuity Bounds\\for Quantum Relative Entropies}

%%%%% Many authors with many affiliations:
\author{%
  \IEEEauthorblockN{Andreas Bluhm\IEEEauthorrefmark{1},
                    {\'A}ngela Capel\IEEEauthorrefmark{2},
                    Paul Gondolf\IEEEauthorrefmark{2},
                    and Antonio Pérez-Hernández\IEEEauthorrefmark{3}}

  \IEEEauthorblockA{\IEEEauthorrefmark{1}%
                    Univ. Grenoble Alpes, CNRS, Grenoble INP, LIG, 38000 Grenoble, France, andreas.bluhm@univ-grenoble-alpes.fr}
  \IEEEauthorblockA{\IEEEauthorrefmark{2}%
                    Fachbereich Mathematik, Univ.  T\"ubingen, 72076 T\"ubingen, Germany,
                    \{paul.gondolf, angela.capel\}@uni-tuebingen.de}
  \IEEEauthorblockA{\IEEEauthorrefmark{3}%
                    Depart. de Matem\'{a}tica Aplicada I,  Univ. Nacional de Educación a Distancia, 28040 Madrid, Spain,
                    antperez@ind.uned.es}
}

\maketitle

%%%%%
%% Abstract: 
%% If your paper is eligible for the student paper award, please add
%% the comment "THIS PAPER IS ELIGIBLE FOR THE STUDENT PAPER
%% AWARD." as a first line in the abstract. 
%% For the final version of the accepted paper, please do not forget
%% to remove this comment!
%%
\begin{abstract}
  In this article, we generalize a proof technique by Alicki, Fannes and Winter and introduce a method to prove continuity bounds for entropic quantities derived from different quantum relative entropies. For the Umegaki relative entropy, we mostly recover known almost optimal bounds, whereas, for the Belavkin-Staszewski relative entropy, our bounds are new. Finally, we use these continuity bounds to derive a new entropic uncertainty relation. This is a short version of \cite{bluhm2022continuity}.
\end{abstract}

\section{Introduction}
\label{sec:intro}

%\ac{Around 1 page and a half of introduction to the problem, brief summary of results, and applications}

Entropic quantities are essential in order to understand processes both in classical and quantum information theory. Examples include (conditional) entropy, (conditional) mutual information, and many others. In applications, one often needs to bound these quantities in terms of a metric distance to nearby states that can be evaluated more easily. Therefore, continuity bounds for entropic quantities have become indispensable tools in quantum information theory. They take the form of a bound on 
\begin{equation*}
    \sup\{|g(\rho) - g(\sigma)|: \rho, \sigma \in \mathcal S_0, d(\rho, \sigma) \leq \epsilon\}.
\end{equation*}
Here, $g$ is the entropic quantity of interest, $\mathcal S_0$ is a suitable subset of the set of quantum states $\cS(\cH)$ in a finite-dimensional Hilbert space, and $d$ is a metric on $\cS(\cH)$.

Bounds of this form have a long history. In 1973, Fannes \cite{Fannes-ContinuityEntropy-1973} proved uniform continuity bounds for the von Neumann entropy, which were sharpened in \cite{Audenaert-ContinuityEstimateEntropy-2007, Petz2008}. Later Alicki and Fannes proved an inequality for the conditional entropy \cite{AlickiFannes-2004}, which Winter improved in \cite{Winter-AlickiFannes-2016} to an almost tight version.

As realized by Shirokov \cite{Shirokov-AFWmethod-2020, Shirokov-ContinuityReview-2022}, the proof by Winter and related versions \cite{mosonyi2011quantum, synak2006asymptotic} do not only work for the conditional entropy but can be generalized and made applicable to a great variety of entropic quantities. Shirokov coined it the \emph{Alicki-Fannes-Winter (AFW) method}. The present article continues along this line of work, further generalising the method. Our aim is to go beyond entropic quantities derived from the Umegaki relative entropy \cite{Umegaki-RelativeEntropy-1962}, defined as
\begin{equation*}
    D(\rho\| \sigma) := \mathrm{Tr}[\rho (\log \rho - \log \sigma)]
\end{equation*}
for quantum states $\rho$, $\sigma$ such that $\ker \sigma \subseteq \ker \rho$ and $+\infty$ otherwise, and to be able to prove continuity bounds for entropic quantities derived from the Belavkin-Staszewski relative entropy (BS-entropy) \cite{BelavkinStaszewski-BSentropy-1982} as well. The \textit{BS-entropy} is defined as 
\begin{equation*}
    \widehat D(\rho\Vert\sigma) := \mathrm{Tr}[\rho \log(\rho^{1/2} \sigma^{-1} \rho^{1/2}) ]
\end{equation*}
for quantum states $\rho$, $\sigma$ such that $\ker \sigma \subseteq \ker \rho$ and $+\infty$ otherwise.

Both Umegaki's relative entropy and BS-entropy reduce to the classical Kullback-Leibler relative entropy \cite{KullbackLeibler-KLD-1951} if the quantum states commute, but the BS-entropy is much less studied \cite{Hiai2017, matsumoto2010reverse, Matsumoto2018}. Recently, there has been renewed interest in the BS-entropy and its properties \cite{Bluhm2020, BluhmCapelPerezHernandez-WeakQFBSentropy-2021} because of its applications to channel capacities \cite{FangFawzi-GeometricRenyiDivergences-2019} and quantum many-body systems \cite{bluhm2022exponential, bardet2021entropy, bardet2021rapid}. In the latter applications, the authors make use of the \textit{BS-mutual information}
\begin{equation*}
    \widehat I_\rho(A:B) := \widehat D(\rho_{AB}\| \rho_A \otimes \rho_B),
\end{equation*}
which is defined as the mutual information, but replacing the Umegaki relative entropy by the BS-entropy. Other entropic quantities derived from the BS-entropy have been considered in \cite{Scalet2021, ZhaiYangXi-BSEntropy-2022}.

In Section \ref{alaff-method}, we introduce our method. First, we choose a quantum relative entropy $\mathbb D(\rho\|\sigma)$, for example, the Umegaki relative entropy. Then, we prove that it is \textit{almost concave}, i.e.,
    \begin{equation*}
        -f(p) \le \mathbb D(\rho_p\|\sigma_p) - p \mathbb D(\rho_0\|\sigma_0) - (1 - p) \mathbb D(\rho_1\|\sigma_1) \le 0
    \end{equation*}
for $\rho_p = p \rho_0 + (1-p) \rho_1$, $\sigma_p = p \sigma_0 + (1-p) \sigma_1$, with $\rho_0, \rho_1, \sigma_0, \sigma_1 \in \mathcal{S}(\mathcal{H})$, $p \in [0,1]$ and a proper function $f$.

\begin{figure}[h!]
    \centering
    \scalebox{0.68}{\begin{tikzpicture}
	\begin{pgfonlayer}{nodelayer}
		\node [style=RoundedOrangeBox] (0) at (-19, -1.5) {Relative entropy\\$\mathbb D(\rho\Vert\sigma)$};
		\node [style=RoundedBlueBox] (1) at (-10, 1) {Convexity\\$\mathbb D(\rho_p\Vert\sigma_p) \le p\mathbb D(\rho_1\Vert\sigma_1) + (1 - p)\mathbb D(\rho_2\Vert\sigma_2)$};
		\node [style=RoundedBlueBox] (2) at (-10, -4) {Almost Concavity\\$\mathbb D(\rho_p\Vert\sigma_p) \ge p\mathbb D(\rho_1\Vert\sigma_1) + (1 - p)\mathbb D(\rho_2\Vert\sigma_2) - f(p)$};
		\node [style=WhiteBox] (3) at (-7, 4.5) {Conditional\\ mutual information\\{\footnotesize$ \mathbb I_\rho(A:B|C) :=  $}\\{\footnotesize$ \mathbb H_\rho(A|C) - \mathbb H_\rho(A|BC) $}};
		\node [style=WhiteBox] (4) at (-18, 6.75) {Mutual information\\{\footnotesize$ \mathbb I_\rho(A:B) :=  $}\\{\footnotesize$  \mathbb D(\rho_{AB} \| \rho_A \otimes \rho_B) $}};
		\node [style=WhiteBox] (5) at (-7, -12) {Relative entropy\\(fixed second argument)\\{\footnotesize$ \left| \mathbb D (\rho_1 \| \sigma ) -  \mathbb D (\rho_2 \| \sigma ) \right| $}\\{\footnotesize$  \leq f_{D,2}(\| \rho_1 - \rho_2 \|_1)$}};
		\node [style=WhiteBox] (6) at (-7, -7.5) {Relative entropy\\(fixed first argument)\\{\footnotesize$ \left| \mathbb D (\rho \| \sigma_1 ) -  \mathbb D (\rho \| \sigma_2 ) \right| $}\\{\footnotesize$  \leq f_{D,1}(\| \sigma_1 - \sigma_2 \|_1)$}};
		\node [style=WhiteBox] (7) at (-18, -12) {Entropy bound\\{\footnotesize$  \mathbb D (\rho \| \sigma )    \leq f_{DB}(\| \rho - \sigma \|_1)$}};
		\node [style=WhiteBox] (11) at (-7, 8.5) {Conditional relative entropy\\{\footnotesize$\mathbb H_\rho(A|B) :=  $}\\{\footnotesize$ - \mathbb D(\rho_{AB} \| \mathds 1_A \otimes \rho_B) $}};
		\node [style=WhiteBox] (13) at (-18, -7.5) {Relative entropy\\{\footnotesize$ \left| \mathbb D (\rho_1 \| \sigma_1 ) -  \mathbb D (\rho_2 \| \sigma_2 ) \right| $}\\{\footnotesize$  \leq f_D(\| \rho_1 - \rho_2 \|_1,\| \sigma_1 - \sigma_2 \|_1)$}};
		\node [style=none] (14) at (-19, 1) {};
		\node [style=none] (15) at (-19, -4) {};
		\node [style=none] (16) at (-1, 1) {};
		\node [style=none] (17) at (-1, -4) {};
		\node [style=none] (18) at (3, -1.5) {};
		\node [style=none] (19) at (3, -7.5) {};
		\node [style=none] (20) at (3, -12) {};
		\node [style=none] (21) at (3, 4.5) {};
		\node [style=none] (22) at (3, 6.75) {};
		\node [style=none] (23) at (3, 8.5) {};
		\node [style=Empty] (24) at (-1, -1.5) {\small Uniform continuity \\\small\& Continuity bounds};
	\end{pgfonlayer}
	\begin{pgfonlayer}{edgelayer}
		\draw [style=thickArrow, in=360, out=180, looseness=0.50] (5) to (7);
		\draw [style=thickArrow, in=0, out=-180] (6) to (13);
		\draw [style=thickLine] (14.center) to (0);
		\draw [style=thickArrow] (14.center) to (1);
		\draw [style=thickLine] (0) to (15.center);
		\draw [style=thickArrow, in=180, out=0] (15.center) to (2);
		\draw [style=thickLine] (1) to (16.center);
		\draw [style=thickLine] (17.center) to (2);
		\draw [style=thickLine] (18.center) to (21.center);
		\draw [style=thickArrow] (22.center) to (4);
		\draw [style=thickLine] (18.center) to (19.center);
		\draw [style=thickLine] (19.center) to (20.center);
		\draw [style=thickArrow] (21.center) to (3);
		\draw [style=thickArrow] (23.center) to (11);
		\draw [style=thickLine] (21.center) to (22.center);
		\draw [style=thickLine] (22.center) to (23.center);
		\draw [style=thickArrow] (19.center) to (6);
		\draw [style=thickArrow] (20.center) to (5);
		\draw [style=thickArrow] (17.center) to (24);
		\draw [style=thickArrow] (16.center) to (24);
		\draw [style=thickLine] (24) to (18.center);
		\draw [style=thickArrow, in=0, out=-180] (5) to (13);
	\end{pgfonlayer}
\end{tikzpicture}}
    \caption{The proof of convexity and concavity of a divergence gives, through our method, a variety of continuity bounds for entropic quantities derived from that divergence. It further allows concluding continuity bounds for the divergence itself. In the above figure, we depicted the scheme and listed quantities that our approach produces results for.}
    \label{fig:concept-Umegaki}
\end{figure}
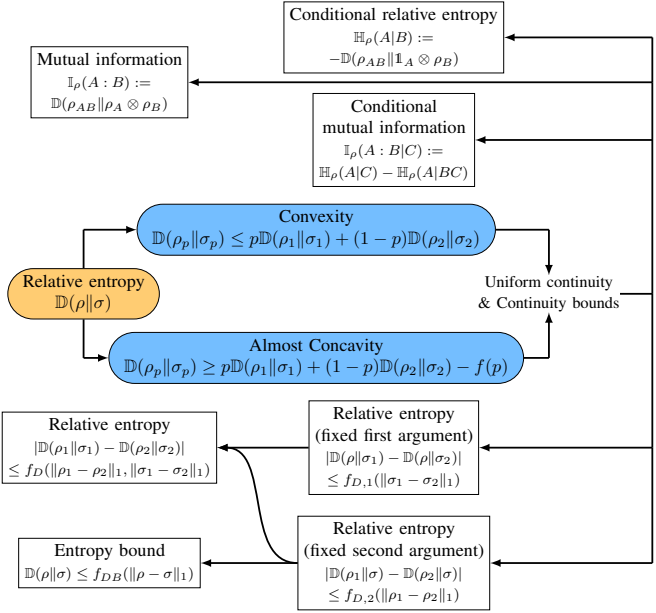

For suitably chosen subsets of the set of quantum states which we call \textit{perturbed $\Delta$-invariant subsets}, the almost concavity of the quantum relative entropy translates into the \textit{almost local affinity} (c.f. Definition~\ref{def:almost-locally-affine}) of derived entropic quantities $g$, i.e.,
    \begin{equation*}
        -a_g(p) \le g(p \rho + (1 - p) \sigma) - p g(\rho) - (1 - p) g(\sigma) \le b_g(p)
    \end{equation*}
for functions $a_g$, $b_g$. That means that the entropic quantity is close to being convex and concave. Our method allows us to convert the almost local affinity into uniform continuity bounds on the entropic quantities $g$, where the continuity bounds are given in terms of how close the entropic quantity is to being convex and concave:
\begin{equation*}
    \sup_{\substack{\rho, \sigma \in \mathcal S_0\\d(\rho, \sigma) \leq \epsilon}}|g(\rho) - g(\sigma)| \leq C(\varepsilon;a_g, b_g).
\end{equation*}
This is illustrated in Figure \ref{fig:concept-Umegaki}. Restriction to perturbed $\Delta$-invariant subsets is vital to the proof of the AFW method and the well-known discontinuities of the Umegaki as well as the Belavkin-Staszewski relative entropies further make it necessary to use the more general perturbed property instead of the original $\Delta$-invariance presented in \cite{Shirokov-ContinuityReview-2022}.
% \vspace{-0.4cm}

While the primary focus lies on the introduction of a method to derive uniform continuity bounds in a unified manner, we showcase its usefulness by first applying it to the Umegaki relative entropy in Section \ref{subsec:Umegaki-entropy} and then to the BS-entropy in Section \ref{subsec:BS-entropy}. We begin by establishing almost concavity of the relative entropies and then present the various continuity bounds that follow from our method for each entropy separately.

We continue with Section \ref{sec:applications}, where we use the obtained bounds to derive results in various contexts within quantum information theory, namely new entropic uncertainty relations, continuity of optimized quantities and robustness of the set of approximate quantum Markov chains. Finally, we conclude in Section \ref{sec:conclusion} with a short discussion and an outlook.

\section{Method}
\label{alaff-method}
%As mentioned in the introduction, 
Our method derives continuity bounds for functions from the quantum states to the reals, such as quantities based on a relative entropy (e.g. conditional entropy, mutual information, and $\rho \mapsto \mathbb{D}(\rho \Vert \sigma)$ with fixed $\sigma$). However, it should be noted that the method is not limited to this specific setting.
%although we use the method in this context it is not limited to this specific setting.

The method is based on two main ingredients. First, the boundedness of the function on a set of quantum states, and second a property called almost local affinity (as defined in Definition~\ref{def:almost-locally-affine}). We mentioned earlier that for quantities related to a relative entropy, almost local affinity can be obtained from the convexity and almost concavity of this relative entropy, provided the almost concave remainder is "well-behaved". This means that under the reduction from the relative entropy to the entropic quantity of choice (possibly also restricting the set of input states), the almost concave bound of the divergence becomes independent of the input state.

Before giving the main theorem, we will introduce two definitions that are necessary for its formalisation. The first is the concept of almost local affinity.

\begin{defn}[Almost locally affine]\label{def:almost-locally-affine}
    Let $f$ be a real-valued function on the convex set $\cS_0 \subseteq \cS(\cH)$. We say that $f$ is \textit{almost locally affine}, if there exist continuous functions $a_f, b_f : [0, 1] \to \R$, which are non decreasing on $[0, \tfrac{1}{2}]$, vanish as $p \to 0^+$ and satisfy
    \begin{equation}\label{eq:alaff-property}
        -a_f(p) \le f(p \rho + (1 - p) \sigma) - p f(\rho) - (1 - p) f(\sigma) \le b_f(p)
    \end{equation}
    for all $p \in [0, 1]$ and $\rho, \sigma \in \cS_0$.
\end{defn}

In the introduction, we mentioned the perturbed $\Delta$-invariant subsets. This property, together with convexity, constitutes a constraint to the domain of the entropic quantity investigated. The property is needed, as it guarantees that for two elements of a set, two interpolation states, or "intermediate states" are as well contained in the set.

\begin{defn}[Perturbed $\Delta$-invariant]
    A set $\cS_0 \subseteq \cS(\cH)$  is called perturbed $\Delta$-invariant with perturbance parameter $t \in [0, 1)$, if for all $\rho, \sigma \in \cS_0$ with $\rho \neq \sigma$, there exists a state $\tau$ such that both states
    \begin{IEEEeqnarray*}{lCr}
        \Delta^\pm(\rho, \sigma, \tau) &= t \tau + (1 - t) \varepsilon^{-1}[\rho - \sigma]_\pm \, ,
    \end{IEEEeqnarray*}
    lie in $\cS_0$. In the above $\varepsilon = \tfrac{1}{2}\norm{\rho - \sigma}_1$, and $[\,\cdot\,]_\pm$ denote the positive and negative parts of a self-adjoint operator, respectively. For $t = 0$, this definition reduces to that in \cite{Shirokov-ContinuityReview-2022}.
\end{defn}

Now that we have introduced all the necessary definitions, we can formulate the method.

\begin{thm}\label{thm:alaff-method}
    Let $\cS_0 \subseteq \cS(\cH)$ be a perturbed $\Delta$-invariant, convex subset of $\cS(\cH)$ containing more than one element, and let $f$ be an almost locally affine (ALAFF) function. If the function $f$ satisfies
    \begin{equation}
        C_f^t := \sup\limits_{\twoline{\rho, \sigma \in \cS_0}{\frac{1}{2}\norm{\rho - \sigma}_1 = 1 - t}} |f(\rho) - f(\sigma)| < + \infty \, .
    \end{equation}
    then it is uniformly continuous, and we have for all $\varepsilon \in (0, 1]$
    \begin{IEEEeqnarray}{lCr}
        \sup\limits_{\twoline{\rho, \sigma \in \cS_0}{\frac{1}{2}\norm{\rho - \sigma}_1 \le \varepsilon}} |f(\rho) - f(\sigma)|\nonumber\\ 
        \label{eq:continuity-bound}
        \le C_f^t \frac{\varepsilon}{1 - t} + \frac{1 - t + \varepsilon}{1 - t} E_f^{\max}\Big(\frac{\varepsilon}{1 - t + \varepsilon}\Big) \, , 
    \end{IEEEeqnarray}
    where $E_f^{\max}:[0, 1) \to \R,$
    \begin{IEEEeqnarray*}{lCr}
     \qquad p \mapsto E_f^{\max}(p) = (1 - p) \max\left\{\frac{E_f(s)}{1 - s} \;:\; 0 \le s \le p\right\} \, .
    \end{IEEEeqnarray*}
    Here we defined $E_f = a_f + b_f$ (c.f. Definition~\ref{def:almost-locally-affine}). Note that for $\varepsilon \in (0, 1 - t]$ $E_f$ and $E_f^{\max}$ coincide.
\end{thm}

\section{Relative Entropies}
\label{sec:relative-entropies}

The natural method to derive the almost local affinity property of entropic quantities relies on having "well-behaved" almost concave bounds for the relative entropy $\mathbb D(\rho\|\sigma)$ they are subordinate to. A divergence $\mathbb D(\rho\|\sigma)$ is called \emph{almost (jointly) concave} on a convex subset $\cS_0 \subset \cS(\cH) \times \cS(\cH)$ if for every $(\rho_{j}, \sigma_{j}) \in \cS_{0}$  ($j=0,1$) there is a function $f(p)$ such that for every $p \in [0,1]$
\begin{equation*}
    -f(p) \le \mathbb D(\rho_p\|\sigma_p) - p \mathbb D(\rho_0\|\sigma_0) - (1 - p) \mathbb D(\rho_1\|\sigma_1) \le 0 \, ,
\end{equation*}
where $\rho_{p} := p \rho_{0} + (1-p) \rho_{1}$ and $\sigma_{p} := p \sigma_{0} + (1-p) \sigma_{1}$. The well-known (joint) convexity  of the Umegaki \cite{WildeFromClassicalToQuantumInformation_2016} and BS \cite{HiaiMosonyi_2011, matsumoto2010reverse} relative entropies ensures that the inequality on the right-hand-side holds. Our main results provide a remainder $f(p)$ involving the function
\begin{equation*}
    f_{\alpha_{1}, \alpha_{2}}(p) = p \log(p+(1-p)\alpha_{1}) + (1-p)\log((1-p)+p\alpha_{2}) \, . 
\end{equation*}
Here, $\alpha_{1}$ and $\alpha_{2}$ are scalar constants defined in terms of the states $\rho_{j}, \sigma_{j}$ via the map
\begin{equation*}
    \alpha(O,P,Q) := \int_{-\infty}^{+\infty} dt \beta_{0}(t) \operatorname{tr}\left[ O P^{\frac{1+it}{2}} Q P^{\frac{1-it}{2}}\right] \, . 
\end{equation*}
In the above $O,P,Q$ are positive semi-definite operators (exponentiation and inverses are restricted to the support), and $\beta_{0}(t)dt$ is a suitable probability measure on $\mathbb{R}$, see \eqref{eq:beta0}. 
\begin{thm}\label{thm:UmegakiAC}
    The Umegaki relative entropy $D(\cdot\|\cdot)$ is almost concave on the subset $\cS_{0} \subset \cS(\cH) \times \cS(\cH)$ of all pairs $(\rho, \sigma)$ such that $\ker \rho \subset \ker \sigma$ with remainder function
    \begin{equation}\label{equa:UmegakiAC}
    f(p) = h(p) \frac{1}{2} \| \rho_{1} - \rho_{2} \|_{1} + f_{c_{1}, c_{2}}(p) \,,
    \end{equation}
    where $c_{1} = \alpha(\rho_{1}, \sigma_{1}^{-1}, \sigma_{2})$ and $c_{2} = \alpha(\rho_{2}, \sigma_{2}^{-1}, \sigma_{1})$.
\end{thm}

As for the BS-relative entropy, we require that the second input state belongs to the set  $\cS_{+}(\cH)$ of full-rank states.

\begin{thm}\label{thm:BSrelAC}
    The BS-relative entropy $\widehat{D}(\cdot\|\cdot)$ is almost concave on $\cS_{0} = \cS (\cH) \times \cS_{+}(\cH)$ with remainder function
    \begin{equation*}
         \widehat{f}(p) = h(p) (1-\delta_{\rho_{1} \rho_{2}}) \widehat{c}_{0} + f_{\widehat{c}_{1}, \widehat{c}_{2}}(p)\,,
    \end{equation*}
    where $\delta_{\rho_{1}, \rho_{2}} \in \{ 0,1\}$ is one if and only if $\rho_{1}=\rho_{2}$, and 
    \begin{IEEEeqnarray*}{ll}
        \widehat{c}_{0} & = \max{\{ \|\sigma_{1}^{-1}\|_{\infty},  \| \sigma_{2}^{-1}\|_{\infty} \}} \,, \\
        \widehat{c}_{1} & = \alpha(\rho_{1}, \rho_{1}^{1/2}\sigma_{1}^{-1} \rho_{1}^{1/2}, \rho_{1}^{-1/2}\sigma_{2}\rho_{1}^{-1/2}) \,, \\
        \widehat{c}_{2} & = \alpha(\rho_{2}, \rho_{2}^{1/2}\sigma_{2}^{-1} \rho_{2}^{1/2}, \rho_{2}^{-1/2}\sigma_{1}\rho_{2}^{-1/2}) \,.
    \end{IEEEeqnarray*}
\end{thm}
The expressions for $f(p)$ and $\widehat{f}(p)$ simplify in some particular cases.\\

\begin{tabular}{ |c|c|c|  }
    \hline
    \emph{Special cases} & Umegaki & BS-entropy \\ \hline \hline
    $p=0,1$  & $f(p)=0$ & $f(p)=0$\\ \hline \\[-3.7mm]
    $\sigma = \sigma_{1} = \sigma_{2}$&   $f(p)=h(p)$ & $\widehat{f}(p)= \widehat{c}_0 h(p)$ \\ \hline 
    $\cH = \cH_{A} \otimes \cH_{B}$ & & \\
    $\sigma_{i} = (\rho_{i})_{A} \otimes \identity_B$ & $f(p) = h(p)$ & $\widehat{f}(p) = \widehat{c}_{0} h(p)$ \\
    $i=1,2$ &   &$ + f_{\hat{c}_0, \hat{c}_0}(p)$\\
    \hline
\end{tabular}

\subsection{Umegaki Relative Entropy}\label{subsec:Umegaki-entropy}

The estimate given in Theorem \ref{thm:UmegakiAC} is tight: if $\dim(\cH) \geq 2$, and we fix two orthonormal pure states $\ket{0},\ket{1}$ in $\cH$ and any $t \in (0,1)$, then the states $\rho_{0} := \ket{0} \bra{0}$, $\rho_{1}:=\ket{1}\bra{1}$, $\sigma_{0}:=t\ket{0}\bra{0}+(1-t)\ket{1}\bra{1}$ and $\sigma_{1}:=(1-t)\ket{0}\bra{0} + t \ket{1} \bra{1}$  satisfy that for every $p \in [0,1]$ 
\[ f(p) = p D(\rho_{0}\Vert\sigma_{0}) + (1-p) D(\rho_{1}\Vert\sigma_{1}) - D(\rho_{p} \Vert \sigma_{p})\,,\]
where $f(p)$ is given in \eqref{equa:UmegakiAC}. 

Combining Theorems \ref{thm:UmegakiAC} and \ref{thm:alaff-method}, we can provide estimates for the uniform continuity of $D(\cdot \Vert \sigma)$ on the set of states $\rho$ with $\ker \sigma \subset \ker \rho$. As a consequence, we obtain that for $\rho, \sigma$ as before, denoting $\varepsilon:=\frac{1}{2}\| \rho - \sigma\|_{1} \leq 1$ and $r(\varepsilon):= (1+\varepsilon) h (\varepsilon/(1+\varepsilon))$ it holds that 
\begin{equation}\label{equa:divergenceBounds} 
    D(\rho \Vert \sigma) \leq \varepsilon \log \widetilde{m}_{\sigma}^{-1} + r(\varepsilon) \leq \left( 1 + \frac{\log \widetilde{m}_{\sigma}^{-1}}{\sqrt{2}}\right) \sqrt{2 \varepsilon} \,,
\end{equation}
where $\widetilde{m}_{\sigma}$ is the minimal non-zero eigenvalue of $\sigma$. Let us remark that alternative estimates of $D(\rho\Vert\sigma)$ have been provided before \cite{AudenaertEisert_II_2011, BraRob81, Vershynina_2019}, all requiring that $\sigma$ has full rank, in contrast to the above result. Moreover, the logarithmic scaling with $\widetilde{m}_{\sigma}^{-1}$ in \eqref{equa:divergenceBounds}  improves the linear dependence that appears in the results by Bratteli-Robinson \cite{BraRob81} and Vershynina \cite{Vershynina_2019}. Numerical simulations show, however, a slight advantage of the estimate by Audenaert-Eisert \cite{AudenaertEisert_II_2011}.

Combining Theorems \ref{thm:BSrelAC} and \ref{thm:alaff-method} we also deduce continuity estimates for well-known entropic quantities derived from the Umegaki relative entropy. Let us recall the definitions and describe the corresponding bounds. For a state on a bipartite system $\rho \in \cS(\cH_{A} \otimes \cH_{B})$ the \emph{conditional entropy} (CE) is given by
\begin{IEEEeqnarray*}{rl}  
    \widehat{H}_{\rho}(A|B) & := -D(\rho_{AB}\Vert\identity_{A} \otimes \rho_{B})\\ 
    & = \max_{\sigma_{B} \in \cS(\cH_{B})} - D(\rho_{AB}\Vert\identity_{A} \otimes \sigma_{B}) \, ,
\end{IEEEeqnarray*}
and the \emph{mutual information} (MI) by
\begin{equation*}
    I_{\rho}(A:B) := D(\rho_{AB}\Vert\rho_{A} \otimes \rho_{B})=S(\rho_{A})-H_{\rho}(A|B) \, .
\end{equation*}
For a state on a tripartite system $\rho \in \cS(\cH_{A} \otimes \cH_{B} \otimes \cH_{C})$ the \emph{conditional mutual information} (CMI) is defined as
\begin{equation*}
     I_{\rho}(A:B|C) = H_{\rho}(A|C) - H_{\rho}(A|BC) \, .
\end{equation*}
Fixed $\rho, \sigma \in \cS(\cH)$ and $\varepsilon$ with $\| \rho - \sigma\|_{1}/2 \leq \varepsilon \leq 1$ and $r(\varepsilon)$ as above, we have the following estimates:
\vspace{-0.25cm}
\begin{flushleft}
    \begin{tabular}{|l|c|} \hline
        \hspace{1.2cm} Entropic Quantity & Upper Bound\\ \hline \hline \\[-3.5mm]
        \begin{tabular}{@{}l@{}}
        CE: $|H_{\rho}(A|B)-H_{\sigma}(A|B)|$\end{tabular} & $2 \varepsilon \log{d_{A}} + r(\varepsilon)$ \\[1mm] \hline
        \begin{tabular}{@{}l@{}}
        MI: $| I_{\rho}(A:B) - I_{\sigma}(A:B)| $ \\[1mm] \hline \\[-3.5mm] CMI: $| I_{\rho}(A:B|C) - I_{\sigma}(A:B|C)| $ \end{tabular}   & \begin{tabular}{@{}c@{}} $2 \varepsilon \, \log \min\{ d_{A}, d_{B}\}$ \\[0.5mm] \hspace{7mm} $+ 2r(\varepsilon)$ \end{tabular}\\ \hline 
    \end{tabular}
\end{flushleft}

Notice that the estimate on the conditional entropy coincides with the result of Winter \cite{Winter-AlickiFannes-2016} which he proved to be almost tight. As for the MI and the CMI, the estimates coincide with the tightest previously known results, see e.g. \cite{Shirokov-AFWmethod-2020} and \cite[Lemma 4]{Shirokov-ContinuityBounds-2019}, respectively. 

Our method also allows us to derive continuity bounds for $D(\rho \| \cdot)$ on the set of states $\sigma$ with ker $ \sigma \subset $ ker $ \rho$ and  $\widetilde{m} \rho \leq \sigma $ for $1>\widetilde{m}>0$.  For such a case, we need the full scope of the method and get a rather involved continuity bound, which can with $\varepsilon = \frac{1}{2}\norm{\sigma_1 - \sigma_2}$ be simplified to the following expression
\begin{equation}\label{eq:cont_bound_rel_ent_second_input}
|D(\rho \| \sigma_1) - D(\rho \| \sigma_2) | \leq \frac{2\varepsilon}{l_{\widetilde m}} \log \widetilde m^{-1} + \log\Big(1 + \frac{\varepsilon}{l_{\widetilde m} + \varepsilon}\frac{1}{\widetilde m}\Big) \, .
\end{equation}
This, jointly with the aforementioned continuity bound for $D(\cdot \| \sigma)$, provides continuity bounds for the relative entropy in the same regimes mentioned above. We can derive, with $\varepsilon = \frac{1}{2}\norm{\rho_1 - \rho_2}$, $\delta = \frac{1}{2}\norm{\sigma_1 - \sigma_2}$ the following
\begin{IEEEeqnarray}{rCl}
    \begin{aligned}
        & |D(\rho_1 \| \sigma_1) - D(\rho_2 \| \sigma_2) | \nonumber \\
        &\hspace{1.5cm}\leq(\sqrt{2} + \log \widetilde{m}^{-1}) \varepsilon^{1/2} +  \frac{3\delta}{l_{\widetilde m}}\log\widetilde m^{-1}\\
        &\hspace{2cm} + 2 \log\Big(1 + \frac{\delta}{l_{\widetilde m} + \delta} \frac{1}{\widetilde m}\Big) \, \nonumber .
    \end{aligned}
\end{IEEEeqnarray}
This is, as far as we know, the first continuity bound for the relative entropy in both inputs existing in the literature.

\subsection{Belavkin-Staszewski Relative Entropy}\label{subsec:BS-entropy}

In contrast to the case of relative entropy, the estimate in Theorem \ref{thm:BSrelAC} is not tight. Indeed, the BS and the Umegaki relative entropies for commuting states coincide. However, $\widehat{f}(p)$ does not reduce to $f(p)$ in this case, since $c_{j} = \hat{c}_{j}$ for $j=1,2$, but $\hat{c}_{0}$ is larger than one if $\mathcal{H} \neq \mathbb{C}$. This is strong evidence for the existence of an improved bound.

Combining Theorems \ref{thm:BSrelAC} and \ref{thm:alaff-method}, we can again prove continuity and divergence bounds for entropic quantities related to the BS-relative entropy. Here again, the full scope of the method is needed, since former quantities are discontinuous for positive semi-definite states. Let us recall the main definitions and some subtle differences with respect to their counterparts. The \emph{BS-conditional entropy} is defined for a state on a bipartite system $\rho \in \cS(\cH_{A} \otimes \cH_{B})$ as
\begin{equation*}
    \widehat{H}_{\rho}(A|B):=-\widehat{D}(\rho_{AB}\Vert\identity_A \otimes \rho_{B} )\,.
\end{equation*}
This quantity is discontinuous on $\cS(\cH)$ and not only do numerics indicate that it differs from the variational definition
\begin{equation*}
    \widehat{H}_{\rho}^{var}(A|B) = \sup_{\sigma_{B} \in \cS(\cH_{B})} - \widehat{D}(\rho_{AB}\Vert\identity_{A} \otimes \sigma_{B}) \, , 
\end{equation*}
where $\widehat{H}_{\rho}(A|B) \leq \widehat{H}_{\rho}^{var}(A|B)$, but we are further able to proof that $\widehat{H}_{\rho}^{var}(A|B)$ is continuous on $\cS(\cH)$ (see Section~\ref{subsec:continuity-of-optimized-quantities}). The \emph{BS-mutual information} is defined as
\begin{equation*}
    \widehat{I}_{\rho}(A:B):= \widehat{D}(\rho_{AB}\Vert\rho_{A} \otimes \rho_{B})
\end{equation*}
and again differs from the variational definition
\begin{equation*}
    \widehat{I}_{\rho}^{var}(A:B):= \inf_{\sigma_{A} \otimes \sigma_{B} \in \cS(\cH_{A} \otimes \cH_{B})}\widehat{D}(\rho_{AB}\Vert\sigma_{A} \otimes \sigma_{B}) \, .
\end{equation*}
Indeed, $\widehat{I}_{\rho}(A:B)$ may scale with $\| \rho_{A}^{-1}\|_{\infty}$ and  $\| \rho_{B}^{-1}\|_{\infty}$, whereas $\widehat{I}_{\rho}^{var}(A:B)$ is bounded by $2 \log \min\{ d_{A}, d_{B}\}$, just like $I_{\rho}(A:B)$. 
For a tripartite  state  $\rho \in \cS(\cH_{A} \otimes \cH_{B} \otimes \cH_{C})$ the \emph{BS-conditional mutual information} is defined by
\begin{equation*}
    \widehat{I}_{\rho}(A:B|C) = \widehat{H}_{\rho}(A|C) - \widehat{H}_{\rho}(A|BC) \, .
\end{equation*}
Fixed $0<m<d_{\cH}^{-1}$ and states $\rho, \sigma$ with minimal eigenvalues $\geq m$ and $\varepsilon:=\frac{1}{2}\| \rho - \sigma\|_{1}$, we have the following estimates:

\begin{flushleft}
\begin{tabular}{|c|c|} \hline
Entropic Quantity & Upper Bound\\ \hline \hline \\[-3.5mm]
\begin{tabular}{@{}l@{}}
 BS-CE:  $|\widehat{H}_{\rho}(A|B)-\widehat{H}_{\sigma}(A|B)|$ \\[1mm] \hline \\[-3.5mm]
BS-MI: $| \widehat{I}_{\rho}(A:B) - \widehat{I}_{\sigma}(A:B)| $ \\[1mm] \hline \\[-3.5mm]
BS-CMI: $| \widehat{I}_{\rho}(A:B|C) - \widehat{I}_{\sigma}(A:B|C)| $ \end{tabular}   & 
\begin{tabular}{@{}c@{}}
$\frac{C \sqrt{\varepsilon}}{m(d_{\mathcal{H}}^{-1} - m)}$ \\[3mm]
{\small ($C:$ absolute }\\[-1mm]
{\small constant)}
\end{tabular} 
\\ \hline 
\end{tabular}
\end{flushleft}
\vspace{3mm}

Although the BS-conditional entropy is discontinuous on the set of positive semi-definite operators, numerics suggest that a bound independent of the minimal eigenvalues of the (full-rank) states might be feasible, or even an almost convexity estimate for the BS-conditional entropy also independent of these minimal eigenvalues. This would also yield an improved bound for the BS-CMI.

For the BS-mutual information, however, we know that a continuity bound will depend on the minimal eigenvalues of the states, since  $\widehat{I}_{\rho}(A:B)$ itself may scale with $\| \rho_{A}^{-1}\|_{\infty}$, as we mentioned before.

\section{Applications}
\label{sec:applications}

The results of almost concavity for the Umegaki and BS relative entropies, as well as the continuity bounds from the previous section, find a plethora of applications in the context of quantum information theory. Here, we show three families of applications: Obtaining new entropic uncertainty relations, deriving continuity of optimized quantities and showing the robustness of the set of approximate quantum Markov chains. 

\subsection{Entropic uncertainty relations}

Let us consider a system $A$ and two POVMs $\mathbf{X}:= \{ X_x\}_{x \in \mathcal{X}}$ and $\mathbf{Y}:= \{ Y_y \}_{y \in \mathcal{Y}}$. Let us further assume the presence of side information $M$ which might help to better predict the outcomes of $\mathbf{X} $ and  $\mathbf{Y} $. Then, for any bipartite state $\rho_{AM} \in \mathcal{S}(\mathcal{H}_A \otimes \mathcal{H}_M)$, the following inequality holds \cite{frank2013extended}:
\begin{IEEEeqnarray}{rCl}\label{eq:entropic_uncertainty_relation}
    H(X | M)_{(\Phi_{\mathbf{X}}\otimes \operatorname{id}_M)(\rho)} + & H(Y | M)_{(\Phi_{\mathbf{Y}}\otimes \operatorname{id}_M)(\rho)} \nonumber \\
    &\geq - \ln c +  H(A | M)_{\rho}  \, ,
\end{IEEEeqnarray}
where $c= \text{max}_{x,y} \{ \operatorname{tr}(X_x Y_y)  \}$ and $\Phi_{\mathbf{Z}}$ is given for $\mathbf{Z} \in \{ \mathbf{X} , \mathbf{Y} \}$ by 
\begin{equation}
    \Phi_{\mathbf{Z}} (\rho_A) := \underset{z \in \mathcal{Z}}{\sum} \operatorname{tr}(\rho_A Z_z) \left| z \right\rangle \left\langle z\right|_Z \, .
\end{equation}
This is known as an \textit{entropic uncertainty relation}. Some references exploring similar inequalities are \cite{berta2010uncertainty} for the special case of measurements in two orthonormal bases and \cite{maasen1988uncertainty} for the case without a memory. We also refer the reader to the review \cite{coles2017entropic} as well as to \cite{gao2018uncertainty} for a more in detail discussion of the setting and related topics. 

In this paper, we consider two von Neumann algebras $\mathcal{N}_X$ and $\mathcal{N}_Y$ such that $\mathcal{N}_X$, resp. $\mathcal{N}_Y$, %is the diagonal onto some orthonormal basis $\big\{ |e_x^{(\mathcal{X})} \rangle \big\}$, resp. $\big\{ |e_y^{(\mathcal{Y})} \rangle \big\}$,
is diagonal in some orthonormal basis $\big\{ |e_x^{(\mathcal{X})} \rangle \big\}$, resp. $\big\{ |e_y^{(\mathcal{Y})} \rangle \big\}$, and we denote by $E_{\mathcal{Z}}$ the Pinching map onto the diagonal spanned by $\big\{|e_z^{(\mathcal{Z})}\rangle \langle e_z^{(\mathcal{Z})}| \big\}$ for $\mathcal{Z} \in \{ \mathcal{X}, \mathcal{Y} \}$. We further consider the algebra $\mathcal{M} = \mathbb{C} \identity_{\ell} \otimes \mathcal{B}(\mathcal{H}_M)$ and the associated map $E_{\mathcal{M}} \otimes \operatorname{id}_M = \frac{1}{d_A} \identity_A \otimes \text{Tr}_A[\cdot]$, for $d_A$ the dimension of $\mathcal{H}_A$. In this setting, a similar inequality to Eq. \eqref{eq:entropic_uncertainty_relation} was proven in \cite{bardet2020approximate}, where $c$ was given by
\begin{equation}
    c:= d_A \, \underset{x,y}{\text{max}}  | \langle e_x^{(\mathcal{X})} | e_y^{(\mathcal{Y})}  \rangle |^2  \, .
\end{equation}

Here, as a consequence of our continuity bound from Eq. \eqref{eq:cont_bound_rel_ent_second_input}, we can prove the following result, which can be viewed as a new entropic uncertainty relation. 

\begin{cor}\label{cor:uncertainty_relations}
    In the conditions above, the following inequality holds for any $\rho_{AM} \in \mathcal{S}(\mathcal{H}_{AM})$:
\begin{IEEEeqnarray}{rCl}
    H(X | M)_{(E_{\mathcal{X}}\otimes \operatorname{id}_M)(\rho_{AM})} + & H(Y | M)_{(E_{\mathcal{Y}}\otimes \operatorname{id}_M)(\rho_{AM})} \nonumber \\
    &\geq - \xi_{\text{RE}} +  H(A | M)_{\rho_{AM}}  \, ,
\end{IEEEeqnarray}
where 
\begin{equation}
    \xi_{\text{RE}}:= \frac{2 \log m^{-1} + m^{-1}}{1 - m}  \, \left(  \underset{x}{\sum} \, \underset{y}{\text{max}} \left| \frac{1}{d_A} -  | \langle e_x^{(\mathcal{X})} | e_y^{(\mathcal{Y})}  \rangle |^2   \right|  \right)^{1/2} \!\!\! , 
\end{equation}
for $m$ given by:
\begin{equation}
     m:= \text{min} \left\{ \frac{1}{(d_A)^2}  , \frac{1}{d_A}\underset{x,y}{\text{min}}  | \langle e_x^{(\mathcal{X})} | e_y^{(\mathcal{Y})}  \rangle |^2 \right\} \, .
\end{equation}
\end{cor}

\subsection{Continuity of optimized quantities}\label{subsec:continuity-of-optimized-quantities}

Given $\mathcal{C} \in \mathcal{S}(\mathcal{H})$ a compact, convex subset of the set of quantum states with at least one positive definite state, we can define the minimal distance to $\mathcal{C}$ in terms of a relative entropy $\mathbb{D}(\cdot \| \cdot)$ as follows:
\begin{equation}
    \mathbb{D}_{\mathcal C}(\rho):= \inf_{\gamma \in \mathcal C} \mathbb{D}(\rho \Vert \gamma).
\end{equation}
The fact that $\mathcal C$ contains a positive definite state guarantees that $D_{\mathcal C}(\rho) < \infty$ for all $\rho \in \mathcal S(\mathcal H)$ (cf. \cite{Winter-AlickiFannes-2016}), and the infimum is attained due to the lower semi-continuity of the relative entropy  \cite{OhyaPetz-Entropy-1993} and Weierstrass’ theorem \cite[Theorem 2.43]{aliprantisborder}. A prominent example of such a $\mathcal C$ includes $\mathrm{SEP}_{AB}$, the set of separable states for systems $A$, $B$.

For $\mathcal C$ as above, we can prove the following results of almost local affinity for the Umegaki relative entropy
\begin{equation*}
    -h(p) \le  D_{\mathcal{C}}( p \rho  + (1-p) \sigma ) - p  D_{\mathcal{C}}(\rho ) - (1 - p)  D_{\mathcal{C}}(\sigma) \le 0 \, ,
\end{equation*}
and for the BS-entropy
\begin{equation*}
    -g_d(p) \le \widehat D_{\mathcal{C}}( p \rho  + (1-p) \sigma ) - p \widehat D_{\mathcal{C}}(\rho ) - (1 - p) \widehat D_{\mathcal{C}}(\sigma) \le 0 \, ,
\end{equation*}
with $g_d(p) := \frac{d}{p^{1/d}}h(p) - \log(1 - p^{1/d})$ for $p \in (0,1)$ and $g_d(0) := 0$. These inequalities, jointly with Theorem \ref{thm:alaff-method}, allow us to prove uniform continuity and derive continuity bounds for any quantity defined as $ D_{\mathcal{C}}( \rho )$ and $ \widehat D_{\mathcal{C}}( \rho )$, with $\mathcal{C}$ as above. In particular, we recover the previously known bound for the relative entropy of entanglement from \cite[Corollary 8]{Winter-AlickiFannes-2016}:
\begin{equation*}
         |D_{\mathrm{SEP}_{AB}}(\rho) - D_{\mathrm{SEP}_{AB}}(\sigma)| \leq \varepsilon \log \min\{d_A,d_B\} +r   \left(\varepsilon  \right) \, .
\end{equation*}
Moreover, we show that $\widehat{H}_{\rho}^{var}(A|B)$ is continuous in $\mathcal{S}(\mathcal{H})$, as opposed to $\widehat{H}_{\rho}(A|B)$, which is only continuous in $\mathcal{S}_{+}(\mathcal{H})$. Following an analogous procedure, we obtain additional continuity bounds for both the Rains and the BS-Rains information. 

\subsection{Approximate Quantum Markov Chains}

Consider a tripartite Hilbert space $\mathcal{H}_{ABC} = \mathcal{H}_A \otimes \mathcal{H}_B \otimes \mathcal{H}_C$,  $\rho_{ABC} \in \mathcal S(\mathcal{H}_{ABC})$ a positive state on it, and the conditional mutual information of $\rho_{ABC}$ between $A$ and $C$ conditioned on $B$. This CMI vanishes if, and only if, it coincides with its Petz recovery  \cite{Petz-MonotonicityRelativeEntropy-2003}
\begin{equation}
\rho_{ABC} = \rho_{AB}^{1/2} \rho_B^{-1/2} \rho_{BC}  \rho_B^{-1/2}  \rho_{AB}^{1/2} \, ,
\end{equation}
i.e., whenever $\rho_{ABC}$ is a quantum Markov chain. As a consequence of our continuity bounds, we can obtain a robustness result for the condition of the quantum Markov chain, namely, we can show that a $\rho_{ABC}$ has a "small" CMI if, and only if, it is close to its Petz recovery. More specifically, combining the continuity bound for the CMI from Section \ref{subsec:Umegaki-entropy} with the findings of  \cite{CarlenVershynina-Stability-DPI-RE-2017}, we have
\begin{IEEEeqnarray}{rCl}
    \begin{aligned}
        &  \left( \frac{\pi}{8} \right)^4 \norm{\rho_{B}^{-1}}_\infty^{-2} \norm{\rho_{ABC}^{-1}}_\infty^{-2}  \norm{\rho_{ABC} - \rho_{AB}^{1/2} \rho_B^{-1/2} \rho_{BC}  \rho_B^{-1/2}  \rho_{AB}^{1/2}}_1^4 \nonumber \\[2mm]
        & \leq I_\rho(A:C | B) \\
        & \leq  2 \left(\log\min\{d_A, d_C\} + 1  \right)  \norm{\rho_{ABC} - \rho_{AB}^{1/2} \rho_B^{-1/2} \rho_{BC}  \rho_B^{-1/2}  \rho_{AB}^{1/2}}_1^{1/2}\, .
    \end{aligned}
\end{IEEEeqnarray}

\section{Conclusion}
\label{sec:conclusion}

In this work, we have introduced a new method to derive uniform continuity bounds for various entropic quantities arising from quantum divergences. We have then applied this method to the Umegaki relative entropy and the BS-entropy. Finally, we have applied the new bounds to prove new entropic uncertainty relations, continuity of optimized quantities and robustness of approximate quantum Markov chains.

While for the Umegaki relative entropy, our bounds recover previous results which are known to be almost tight, there is evidence that our bounds for the entropic quantities arising from the BS-entropy can be improved. First, if $\rho$ and $\sigma$ commute, Umegaki and BS-relative entropy coincide, and so should their continuity bounds. However, our bounds do not have this property. Moreover, numerics suggest that the bound for the BS-conditional entropy should be independent of the minimal eigenvalues of the state, such that our bounds can likely be improved. On the other hand, the BS-conditional entropy exhibits discontinuities for singular states, such that possible eigenvalue-independent continuity bounds can only hold on the set of full-rank states. This behaviour is different from the behaviour of the usual conditional entropy. It is therefore important to better understand the pathologies arising from the definition of the BS-entropy compared to the Umegaki relative entropy. We leave these improvements for future work.

Our method is not limited to the relative entropies studied in this article but works more generally for all almost concave quantum divergences. Therefore, it is natural to ask whether this method can work for quantities such as Tsallis entropies, different variants of R{\'e}nyi entropies (Petz \cite{OhyaPetz-Entropy-1993}, sandwiched \cite{M_ller_Lennert_OnQuantumRenyiEntropies_2013}, geometric \cite{FangFawzi-GeometricRenyiDivergences-2019}), and many more. We are exploring these directions in a subsequent project.

%%%%%%
%% To balance the columns at the last page of the paper use this
%% command somewhere at the top of the first column of the last page:
%%
%%
%% where the exact amount of page reduction has to be adapted to the
%% actual situation.
%%
%% If the balancing should occur in the middle of the references, use
%% the following trigger:
%%
% \IEEEtriggeratref{3}
%%
%% which triggers a \newpage (i.e., new column) just before the given
%% reference number. Note that you need to adapt this if you modify
%% the paper. The "triggered" command can be changed if desired:
%%
% \IEEEtriggercmd{\enlargethispage{-20cm}}
%%
%%%%%%
\section*{Acknowledgment}
The authors are grateful to Li Gao and Mark Wilde for interesting discussions and to Peter Brown for the example of discontinuity of the BS-CE. Moreover, the authors thank {\'A}lvaro Alhambra for mentioning \cite{gour2022role}. The authors would also like to thank L. Lami and M. Tomamichel for spotting an error in an earlier version of the paper. A.P.H acknowledges fin. sup. from the Spanish Ministerio de Ciencia e Innovación (grant PID2020-113523GB-I00), Comunidad de Madrid (grant QUITEMAD-CMS2018/TCS-4342) and ETSI Industriales (UNED) of Spain project 2023-ETSII-UNED-01. This work was partially funded by the Deutsche Forschungsgemeinschaft (DFG, German Research Foundation) – Project-ID 470903074 – TRR 352.

%%%%%%
%% References:
%% We recommend the usage of BibTeX:
%%
\bibliographystyle{IEEEtran}
\bibliography{lit}

\begin{thebibliography}{10}
\providecommand{\url}[1]{#1}
\csname url@samestyle\endcsname
\providecommand{\newblock}{\relax}
\providecommand{\bibinfo}[2]{#2}
\providecommand{\BIBentrySTDinterwordspacing}{\spaceskip=0pt\relax}
\providecommand{\BIBentryALTinterwordstretchfactor}{4}
\providecommand{\BIBentryALTinterwordspacing}{\spaceskip=\fontdimen2\font plus
\BIBentryALTinterwordstretchfactor\fontdimen3\font minus
  \fontdimen4\font\relax}
\providecommand{\BIBforeignlanguage}[2]{{%
\expandafter\ifx\csname l@#1\endcsname\relax
\typeout{** WARNING: IEEEtran.bst: No hyphenation pattern has been}%
\typeout{** loaded for the language `#1'. Using the pattern for}%
\typeout{** the default language instead.}%
\else
\language=\csname l@#1\endcsname
\fi
#2}}
\providecommand{\BIBdecl}{\relax}
\BIBdecl

\bibitem{bluhm2022continuity}
A.~Bluhm, Ángela Capel, P.~Gondolf, and A.~Pérez-Hernández, ``Continuity of
  quantum entropic quantities via almost convexity,'' \emph{arXiv preprint,
  arXiv:2208.00922}, 2022.

\bibitem{Fannes-ContinuityEntropy-1973}
M.~Fannes, ``A continuity property of the entropy density for spin lattice
  systems,'' \emph{Commun. Math. Phys.}, vol.~31, pp. 291--294, 1973.

\bibitem{Audenaert-ContinuityEstimateEntropy-2007}
K.~M.~R. Audenaert, ``A sharp continuity estimate for the {von Neumann}
  entropy,'' \emph{J. Phys. A: Math. Theor.}, vol.~40, no.~28, p. 8127, 2007.

\bibitem{Petz2008}
D.~Petz, \emph{Quantum Information Theory and Quantum Statistics}.\hskip 1em
  plus 0.5em minus 0.4em\relax Springer, 2008.

\bibitem{AlickiFannes-2004}
R.~Alicki and M.~Fannes, ``Continuity of quantum conditional information,''
  \emph{J. Phys. A: Math. Gen.}, vol.~37, no.~5, pp. L55--L57, 2004.

\bibitem{Winter-AlickiFannes-2016}
A.~Winter, ``Tight uniform continuity bounds for quantum entropies: conditional
  entropy, relative entropy distance and energy constraints,'' \emph{Commun.
  Math. Phys.}, vol. 347, pp. 291--313, 2016.

\bibitem{Shirokov-AFWmethod-2020}
M.~E. Shirokov, ``Advanced {Alicki-Fannes-Winter} method for energy-constrained
  quantum systems and its use,'' \emph{Quantum Inf. Process.}, vol.~19, p. 164,
  2020.

\bibitem{Shirokov-ContinuityReview-2022}
------, ``Continuity of characteristics of composite quantum systems: a
  review,'' \emph{arXiv preprint: arXiv 2201.11477}, 2022.

\bibitem{mosonyi2011quantum}
M.~Mosonyi and F.~Hiai, ``On the quantum {R{\'e}nyi} relative entropies and
  related capacity formulas,'' \emph{IEEE Trans. Inf. Theory}, vol.~57, no.~4,
  pp. 2474--2487, 2011.

\bibitem{synak2006asymptotic}
B.~Synak-Radtke and M.~Horodecki, ``On asymptotic continuity of functions of
  quantum states,'' \emph{J. Phys. A: Math. Gen.}, vol.~39, no.~26, p. L423,
  2006.

\bibitem{Umegaki-RelativeEntropy-1962}
H.~Umegaki, ``Conditional expectation in an operator algebra {IV. E}ntropy and
  information,'' \emph{Kodai Math. Sem. Rep.}, vol.~14, pp. 59--85, 1962.

\bibitem{BelavkinStaszewski-BSentropy-1982}
V.~P. Belavkin and P.~Staszewski, ``{$C^*$}-algebraic generalization of
  relative entropy and entropy,'' \emph{Ann. Inst. Henri Poincaré, section A},
  vol.~37, no.~1, pp. 51--58, 1982.

\bibitem{KullbackLeibler-KLD-1951}
S.~Kullback and R.~A. Leibler, ``On information and sufficiency,'' \emph{Annals
  of Math. Stat.}, vol.~22, no.~1, pp. 79--86, 1951.

\bibitem{Hiai2017}
F.~Hiai and M.~Mosonyi, ``Different quantum $f$-divergencies and the
  reversibility of quantum operations,'' \emph{Rev. Math. Phys.}, vol.~29,
  no.~7, p. 1750023, 2017.

\bibitem{matsumoto2010reverse}
K.~Matsumoto, ``Reverse test and characterization of quantum relative
  entropy,'' \emph{arXiv preprint arXiv:1010.1030}, 2010.

\bibitem{Matsumoto2018}
------, ``A new quantum version of $f$-divergence,'' in \emph{Reality and
  Measurement in Algebraic Quantum Theory}.\hskip 1em plus 0.5em minus
  0.4em\relax Springer, 2018, pp. 229--273.

\bibitem{Bluhm2020}
A.~Bluhm and {\'A}.~Capel, ``{A strengthened data processing inequality for the
  Belavkin-Staszewski relative entropy},'' \emph{Rev. Math. Phys.}, vol.~32,
  no.~2, p. 2050005, 2020.

\bibitem{BluhmCapelPerezHernandez-WeakQFBSentropy-2021}
A.~Bluhm, A.~Capel, and A.~Pérez-Hernández, ``{Weak Quasi-Factorization for
  the Belavkin-Staszewski relative entropy},'' \emph{2021 IEEE International
  Symposium on Information Theory (ISIT)}, pp. 118--123, 2021.

\bibitem{FangFawzi-GeometricRenyiDivergences-2019}
K.~Fang and H.~Fawzi, ``Geometric {Rényi} divergence and its applications in
  quantum channel capacities,'' \emph{Commun. Math. Phys.}, vol. 384, pp.
  1615--1677, 2021.

\bibitem{bluhm2022exponential}
A.~Bluhm, {\'A}.~Capel, and A.~P{\'e}rez-Hern{\'a}ndez, ``Exponential decay of
  mutual information for {Gibbs} states of local {Hamiltonians},''
  \emph{Quantum}, vol.~6, p. 650, 2022.

\bibitem{bardet2021entropy}
I.~Bardet, {\'A}.~Capel, L.~Gao, A.~Lucia, D.~P{\'e}rez-Garc{\'\i}a, and
  C.~Rouz{\'e}, ``Entropy decay for {Davies} semigroups of a one dimensional
  quantum lattice,'' \emph{arXiv preprint arXiv:2112.00601}, 2021.

\bibitem{bardet2021rapid}
------, ``Rapid thermalization of spin chain commuting {Hamiltonians},''
  \emph{Phys. Rev. Lett.}, vol. 130, p. 060401, 2023.

\bibitem{Scalet2021}
S.~O. Scalet, A.~M. Alhambra, G.~Styliaris, and J.~I. Cirac, ``Computable
  {Rényi} mutual information: Area laws and correlations,'' \emph{Quantum},
  vol.~5, p. 541, 2021.

\bibitem{ZhaiYangXi-BSEntropy-2022}
Y.~Zhai, B.~Yang, and Z.~Xi, ``{Belavkin–Staszewski} relative entropy,
  conditional entropy, and mutual information,'' \emph{Entropy}, vol.~24, p.
  837, 2022.

\bibitem{WildeFromClassicalToQuantumInformation_2016}
M.~M. Wilde, \emph{Quantum Information Theory}.\hskip 1em plus 0.5em minus
  0.4em\relax Cambridge University Press, 2016.

\bibitem{HiaiMosonyi_2011}
F.~Hiai, M.~Mosonyi, D.~Petz, and C.~B{\'{e}}ny, ``{Quantum} $f$-divergences
  and error correction,'' \emph{Rev. Mat. Phys.}, vol.~23, no.~07, pp.
  691--747, 2011.

\bibitem{AudenaertEisert_II_2011}
K.~M.~R. Audenaert and J.~Eisert, ``Continuity bounds on the quantum relative
  entropy {\textemdash} {II},'' \emph{J. Math. Phys.}, vol.~52, p. 112201,
  2011.

\bibitem{BraRob81}
O.~Bratteli and D.~W. Robinson, \emph{Operator algebras and quantum-statistical
  mechanics II. Equilibrium states. Models in quantum statistical mechanics},
  ser. Texts and Monographs in Physics.\hskip 1em plus 0.5em minus 0.4em\relax
  Springer, 1981.

\bibitem{Vershynina_2019}
A.~Vershynina, ``Upper continuity bound on the quantum quasi-relative
  entropy,'' \emph{J. Math. Phys.}, vol.~60, p. 102201, 2019.

\bibitem{Shirokov-ContinuityBounds-2019}
M.~E. Shirokov, ``Uniform continuity bounds for information characteristics of
  quantum channels depending on input dimension and on input energy,'' \emph{J.
  Phys. A: Math. Theor.}, vol.~52, no.~1, p. 014001, 2019.

\bibitem{frank2013extended}
R.~L. Frank and E.~H. Lieb, ``Extended quantum conditional entropy and quantum
  uncertainty inequalities,'' \emph{Comm. Math. Phys.}, vol. 323, no.~2, pp.
  487--495, 2013.

\bibitem{berta2010uncertainty}
M.~Berta, M.~Christandl, R.~Colbeck, J.~M. Renes, and R.~Renner, ``The
  uncertainty principle in the presence of quantum memory,'' \emph{Nature
  Physics}, vol.~6, no.~9, pp. 659--662, 2010.

\bibitem{maasen1988uncertainty}
H.~Maasen and J.~B.~M. Uffink, ``Generalized entropic uncertainty relations,''
  \emph{Physical Review Letters}, vol.~60, pp. 1103--1106, 1988.

\bibitem{coles2017entropic}
P.~J. Coles, M.~Berta, M.~Tomamichel, and S.~Wehner, ``Entropic uncertainty
  relations and their applications,'' \emph{Rev. Mod. Phys.}, vol.~89, no.~1,
  p. 015002, 2017.

\bibitem{gao2018uncertainty}
L.~Gao, M.~Junge, and N.~LaRacuente, ``Uncertainty principle for quantum
  channels,'' in \emph{2018 IEEE International Symposium on Information Theory
  (ISIT)}.\hskip 1em plus 0.5em minus 0.4em\relax IEEE, 2018, pp. 996--1000.

\bibitem{bardet2020approximate}
I.~Bardet, {\'{A}}.~Capel, and C.~Rouz{\'{e}}, ``Approximate tensorization of
  the relative entropy for noncommuting conditional expectations,'' \emph{Ann.
  Henri Poincar{\'{e}}}, vol.~23, pp. 101--140, Jul. 2022.

\bibitem{OhyaPetz-Entropy-1993}
M.~Ohya and D.~Petz, \emph{Quantum Entropy and Its Use}, ser. Texts and
  Monographs in Physics.\hskip 1em plus 0.5em minus 0.4em\relax Springer, 1993.

\bibitem{aliprantisborder}
C.~D. Aliprantis and K.~C. Border, \emph{Infinite Dimensional Analysis: A
  Hitchhiker's Guide}, 3rd~ed.\hskip 1em plus 0.5em minus 0.4em\relax Springer,
  2006.

\bibitem{Petz-MonotonicityRelativeEntropy-2003}
D.~Petz, ``Monotonicity of quantum relative entropy revisited,'' \emph{Rev.
  Math. Phys.}, vol.~15, no.~1, pp. 79--91, 2003.

\bibitem{CarlenVershynina-Stability-DPI-RE-2017}
E.~A. Carlen and A.~Vershynina, ``Recovery map stability for the data
  processing inequality,'' \emph{J. Phys. A: Math. Theor.}, vol.~53, no.~3, p.
  035204, 2020.

\bibitem{M_ller_Lennert_OnQuantumRenyiEntropies_2013}
M.~Müller-Lennert, F.~Dupuis, O.~Szehr, S.~Fehr, and M.~Tomamichel, ``On
  quantum {R{\'{e}}nyi} entropies: A new generalization and some properties,''
  \emph{J. Math. Phys.}, vol.~54, no.~12, p. 122203, dec 2013.

\bibitem{gour2022role}
G.~Gour, ``On the role of quantum coherence in thermodynamics,'' \emph{arXiv
  preprint arXiv:2205.13612}, 2022.

\bibitem{audenaert2014quantum}
K.~M.~R. Audenaert, ``Quantum skew divergence,'' \emph{J. Math. Phys.},
  vol.~55, p. 112202, 2014.

\bibitem{AndoHiai_PeierlsBogolubov_1998}
T.~Ando and F.~Hiai, ``Hölder type inequalities for matrices,'' \emph{Math.
  Inequal. Appl.}, vol.~1, no.~1, pp. 1--30, 1998.

\bibitem{Araki_LiebThirring_1990}
H.~Araki, ``On an inequality of {Lieb} and {Thirring},'' \emph{Lett. Math.
  Phys.}, vol.~19, pp. 167--170, 1990.

\bibitem{LiebThirring_1976}
E.~H. Lieb and W.~E. Thirring, ``Inequalities for the moments of the
  eigenvalues of the {Schrödinger Hamiltonian} and their relation to {Sobolev}
  inequalities,'' in \emph{Studies in Mathematical Physics}, E.~H. Lieb,
  Ed.\hskip 1em plus 0.5em minus 0.4em\relax Princeton University Press, 1976,
  pp. 269--304.

\bibitem{SutterBertaTomamichel-Multivariate-2017}
D.~Sutter, M.~Berta, and M.~Tomamichel, ``Multivariate trace inequalities,''
  \emph{Commun. Math. Phys.}, vol. 352, pp. 37--58, 2017.

\bibitem{Carlen-TraceInequalities-2009}
E.~A. Carlen, ``Trace inequalities and quantum entropy: An introductory
  course,'' in \emph{Entropy and the Quantum}, ser. Contemporary Mathematics,
  R.~Sims and D.~Ueltschi, Eds.\hskip 1em plus 0.5em minus 0.4em\relax American
  Mathematical Society, 2009, vol. 529, pp. 73--140.

\bibitem{junge2019stability}
M.~Junge, N.~LaRacuente, and C.~Rouz{\'e}, ``Stability of logarithmic {S}obolev
  inequalities under a noncommutative change of measure,'' \emph{J. Stat.
  Phys.}, vol. 190, p.~30, 2023.

\end{thebibliography}

\appendix

\subsection{Proof of Theorem~\ref{thm:alaff-method}}

Let $t \in [0,1)$ and $\varepsilon \in (0, 1]$. Let further $\rho, \sigma \in \cS_0$ with $\frac{1}{2}\norm{\rho - \sigma}_1 = \varepsilon$. By the property of perturbed $\Delta$-invariance there exists at least one $\tau \in \cS(\cH)$ such that $\gamma_\pm := \Delta^\pm(\rho, \sigma, \tau) \in \cS_0$. Now for every such $\gamma_\pm$ we have that
\begin{IEEEeqnarray*}{lr}
\begin{aligned}
    \omega &=\frac{1 - t}{1 - t + \varepsilon}\rho + \frac{\varepsilon}{1 - t + \varepsilon} \gamma_-  \\
    &=\frac{1 - t}{1 - t + \varepsilon} \sigma + \frac{\varepsilon}{1 - t + \varepsilon} \gamma_+ \, ,
  \end{aligned}
\end{IEEEeqnarray*}
which can be easily checked by inserting the explicit form of $\gamma_\pm$ and using that $[\rho - \sigma]_+ - [\rho - \sigma]_- = \rho - \sigma$. We get that $\omega \in \cS_0$, since $\cS_0$ is convex, which allows us to evaluate $f$ at $\omega$ and use Eq.~\eqref{eq:alaff-property} for both of the representations of $\omega$. This gives us
\begin{IEEEeqnarray*}{rCl}
    -a_f(p) &\le f(\omega) - (1 - p) f(\rho) - p f(\gamma_-) \le b_f(p) \, ,\\
    -a_f(p) &\le f(\omega) - (1 - p) f(\sigma) - p f(\gamma_+) \le b_f(p)\, ,
\end{IEEEeqnarray*}
where we set $p = p(\varepsilon) = \frac{\varepsilon}{1 - t + \varepsilon}$ for better readability. Note that $p \in (0, \frac{1}{2 - t}] \subseteq [0, 1)$ as $\varepsilon \in (0, 1]$ and $t \in [0, 1)$ and further that $p(\varepsilon)$ is monotone with respect to $\varepsilon$. We recombine the above to get
\begin{IEEEeqnarray*}{rCl}
    (1 - p) (f(\rho) - f(\sigma)) &\le p (f(\gamma_+) - f(\gamma_-)) + a_f(p) + b_f(p) \, ,\\
    (1 - p) (f(\sigma) - f(\rho)) &\le p (f(\gamma_-) - f(\gamma_+)) + a_f(p) + b_f(p) \, .
\end{IEEEeqnarray*}
Those two inequalities immediately give us
\begin{equation*}
    (1 - p) |f(\rho) - f(\sigma)| \le p |f(\gamma_+) - f(\gamma_-)| + (a_f + b_f)(p) \, .
\end{equation*}
If we now insert $E_f = a_f + b_f$, we obtain
\begin{equation*}
    |f(\rho) - f(\sigma)| \le \frac{p}{1 - p} |f(\gamma_+) - f(\gamma_-)| + \frac{1}{1 - p} E_f(p) \, .
\end{equation*}

In the case that $C_f^t$ is finite, we have $|f(\gamma_+) - f(\gamma_-)| \le C_f^t$ as $\gamma_\pm \in \cS_0$ and fulfil $\frac{1}{2}\norm{\gamma_+ - \gamma_-}_1 = 1 - t$. This allows us to take the supremum over all $\rho$, $\sigma \in \cS_0$ with $\frac{1}{2}\norm{\rho - \sigma}_1 = \varepsilon$ and even extend to $\frac{1}{2}\norm{\rho - \sigma}_1 \le \varepsilon$ in two steps. In the first step we bound $\frac{1}{1 - p}E_f(p)$ from above with $\frac{1}{1 - p}E_f^{\max}(p)$. In the second step, we use that $\frac{1}{1 - p} E_f^{\max}(p)$ is engineered to be non-decreasing on $[0, 1)$. Hence for the specific $p = p(\varepsilon) = \frac{\varepsilon}{1 - t + \varepsilon} \in [0, \frac{1}{2 - t}] \subset [0, 1)$, it is non-decreasing in $\varepsilon$ as well. This gives us the upper bound in Eq.~\eqref{eq:continuity-bound} at last. The reduction of $E_f^{\max}$ to $E_f$ on $\varepsilon \in (0, 1 - t]$ $E_f^{\max}$ is due to $E_f$ being non-decreasing on $[0, \frac{1}{2}]$. Hence, $E_f^{\max}$ inherits the vanishing property as $p \to 0^+$, which translates to $E_f^{\max}(p(\varepsilon)) \to 0$ if $\varepsilon \to 0^+$. Thus we conclude uniform continuity. 

\hfill\IEEEQED

\subsection{Proof of Theorem \ref{thm:UmegakiAC}}

It is clear that $\cS_{0}$ is a convex set and that the bound holds trivially for $p = 0$ and $p = 1$. Hence let $p \in (0, 1)$ and $(\rho_1, \sigma_1), (\rho_2, \sigma_2) \in \cS_{0}$ in the following. We find that
\begin{IEEEeqnarray}{rCl}
    \begin{aligned}
        & p D(\rho_1 \Vert \sigma_1) + (1 - p) D(\rho_2 \Vert \sigma_2) - D(\rho \Vert \sigma)\\ 
        &= - p S(\rho_1) - (1 - p) S(\rho_2) + S(\rho)\\
        &\hspace{1cm} + (1 - p) \tr[\rho_2(\log\sigma -
            \log\sigma_2)]\\
        &\hspace{2cm} + p \tr[\rho_1(\log\sigma - \log\sigma_1)]\\
        &\le h(p)\frac{1}{2}\norm{\rho_1 - \rho_2}_1 + f_{c_1, c_2}(p) \, ,
    \end{aligned}
\end{IEEEeqnarray}
where we split the relative entropies and used that the von Neumann entropy fulfils \cite[Theorem 14]{audenaert2014quantum}
\begin{equation}\label{eq:eq_almost_convexity_von_neumann_entropy}
    S(\rho) \le \frac{1}{2}\norm{\rho_1 - \rho_2}_1 h(p) + p S(\rho_1) + (1 - p) S(\rho_2) \, . 
\end{equation} 
Furthermore, we upper bound the remaining terms by $f_{c_1, c_2}(p)$, estimating the two separately. We will only demonstrate the derivation for the second term, as it is completely analogous to the first one. We have 
\begin{IEEEeqnarray}{rCl}
    \begin{aligned}\label{eq:eq_int_ineq}
        & p\tr[\rho_1 (\log(\sigma) - \log(\sigma_1))]\\ & = p\tr\left[\exp(\log(\rho_1))(\log(\sigma) - \log(\sigma_1))\right]\\
        &\le p \log\tr\left[\exp\left(\log(\rho_1) + \log(\sigma) - \log(\sigma_1)\right)\right]\\
        &\le p \log\int\limits_{-\infty}^\infty dt\,\beta_0(t)\,\tr\left[\rho_1 (\sigma_1^{-1})^{\frac{ 1+it}{2}} \sigma (\sigma_1^{-1})^{\frac{1-it}{2}}\right] \, .
    \end{aligned}  
\end{IEEEeqnarray}
The first estimate follows immediately using the well-known Peierls-Bogolubov inequality \cite{AndoHiai_PeierlsBogolubov_1998}. The second one involves a generalisation of the Araki-Lieb-Thirring inequality \cite{Araki_LiebThirring_1990, LiebThirring_1976} by Sutter et al. \cite[Corollary 3.3]{SutterBertaTomamichel-Multivariate-2017} with
\begin{equation}\label{eq:beta0}
    \beta_0(t) = \frac{\pi}{2}\frac{1}{\cosh(\pi t) + 1}
\end{equation}
a probability density on $\R$. In the above steps, i.e. \eqref{eq:eq_int_ineq}, we relied on $\rho_1, \sigma_1$ and $\sigma$ to be full rank. If this is not the case one can also rigorously obtain the same result taking into account the supporting subspaces, however, the procedure is more involved. We restrict to the full-rank setting for the sake of clarity. Note here that in the most general case $\cdot^{-1}$ in the RHS of \eqref{eq:eq_int_ineq} is the Moore-Penrose pseudoinverse. The trace in the integral can now be estimated for each $t$ by
\begin{IEEEeqnarray}{rCl}
    \begin{aligned}\label{eq:eq_splitting_of_trace}
        \tr[\rho_1 & (\sigma_1^{-1})^{\frac{1+it }{2}} \sigma  (\sigma_1^{-1})^{\frac{1-it}{2}}]\\ & = p + (1 - p)\tr[\rho_1 (\sigma_1^{-1})^{\frac{1+it}{2}} \sigma_2 (\sigma_1^{-1})^{\frac{1-it}{2}}] \, . 
    \end{aligned}
\end{IEEEeqnarray}
Here, we just split $\sigma$ and used the cyclicity of the trace to get rid of the unitary. To see that $c_1 < \infty$, we upper bound $\sigma_2$ by $\identity$ and $\sigma_1^{-1}$ by $\widetilde{m}_{\sigma_1}^{-1}\identity$ where $\widetilde{m}_{\sigma_1}$ is the smallest non-zero eigenvalue of $\sigma_1$. This can be done, since $\ker \sigma_1 \subseteq \ker \rho_1$. We end up with $c_1 \le \widetilde{m}_{\sigma_1}^{-1} < \infty$. Inserting \eqref{eq:eq_splitting_of_trace} into \eqref{eq:eq_int_ineq}, we obtain the first part of $f_{c_1, c_2}(p)$ and repeating the steps for $(1 - p)\tr[\rho_2(\log(\sigma) - \log(\sigma_2))]$ the second one as well. This concludes the proof.

\hfill\IEEEQED

\subsection{Proof of Theorem \ref{thm:BSrelAC}}

We next give some auxiliary results that will be needed later. The first of those concerns an operator inequality for the term inside the trace in the definition of the BS-entropy. 

\begin{lem}\label{lem:almost_concavity_xlogx} Let $A_1, A_2 \in \cB(\cH)$ positive semi-definite, $p \in [0, 1]$ and $A := p A_1 + (1 - p) A_2$. Then
    \begin{IEEEeqnarray*}{rCl}
        -A \log(A) \le -p A_1 \log(A_1) & - (1 - p) A_2 \log(A_2)\\ 
        & + h_{A_1, A_2}(p)\identity
    \end{IEEEeqnarray*}
    with $h_{A_1, A_2}(p) = -p \log(p)\tr[A_1] - (1 - p) \log(1 - p)\tr[A_2]$ a distorted binary entropy.
\end{lem}
\subsubsection*{Proof}
It holds that 
\begin{IEEEeqnarray}{l}
    -A  \log(A) + p A_1 \log(A_1) +  (1 - p) A_2 \log(A_2) \label{eq:eq_norm_upper_bound}\\
    \le \norm{ -A\log(A)  + p A_1 \log(A_1) +  (1 - p) A_2 \log(A_2) }_1 \identity  \, . \nonumber
\end{IEEEeqnarray}
Now, since $x \mapsto -x\log(x)$ is operator concave \cite[Theorem 2.6]{Carlen-TraceInequalities-2009}, we have 
\begin{equation*}
    -A\log(A) \ge -p A_1 \log(A_1) - (1 - p) A_2 \log(A_2) \, ,
\end{equation*}
giving us that 
\begin{equation*}
    -A\log(A) + p A_1 \log(A_1) +  (1 - p) A_2 \log(A_2) \ge 0 \, ,
\end{equation*}
and hence
\begin{IEEEeqnarray}{l}
    \label{eq:eq_almost_convexity_xlogx}
        \norm{-A\log(A) + p A_1 \log(A_1) +  (1 - p) A_2 \log(A_2)}_1\\
        = \tr[-A\log(A) + p A_1 \log(A_1) +  (1 - p) A_2 \log(A_2)] \nonumber\, .
\end{IEEEeqnarray}
We now use the operator monotonicity of the logarithm to find
\begin{IEEEeqnarray*}{rCl}
    & -\tr[A \log(A)] = -p\tr[A_1 \log(A)] - (1 - p) \tr[A_2 \log(A)]\\
    &  \le -p \tr[A_1 \log(p A_1)] - (1 - p) \tr[A_2 \log((1 - p) A_2)]\\
    &  = -p \tr[A_1 \log(A_1)] - (1 - p) \tr[A_2 \log(A_2)] + h_{A_1, A_2}(p)
\end{IEEEeqnarray*}
Inserting this into \eqref{eq:eq_almost_convexity_xlogx} and then into \eqref{eq:eq_norm_upper_bound} yields the claimed result.

\hfill\IEEEQED

The next auxiliary result concerns an equivalent formulation for the BS-entropy constructed from the function $x \mapsto x \log x$.

\begin{lem}\cite[Eq. (7.35)]{OhyaPetz-Entropy-1993}\label{lem:alternative_representation_bs_entropy} Let $\rho \in \cS(\cH)$ and $\sigma \in \cS_+(\cH)$, then
    \begin{equation}
        \widehat{D}(\rho \Vert \sigma) = \tr[\sigma(\sigma^{-1/2} \rho \sigma^{-1/2}) \log(\sigma^{-1/2} \rho \sigma^{-1/2})].
    \end{equation}
\end{lem}

Building on the previous results from this section, we proceed to prove the main result, namely the almost concavity for the BS-entropy.  Let $(\rho_1, \sigma_1), (\rho_2, \sigma_2) \in \cS(\cH) \times \cS(\cH)_{+}$, $p \in [0,1]$ and consider $\rho:=p \rho_{1} + (1-p) \rho_{2}, \sigma:=p \sigma_{1} + (1-p) \sigma_{2}$. The formula for $p = 0, 1$ is trivial, hence let $p \in (0, 1)$. We find that
\begin{IEEEeqnarray}{rCl}
    \begin{aligned}
        p \widehat{D}(\rho_1 \Vert \sigma_1) & + (1 - p) \widehat{D}(\rho_2 \Vert \sigma_2) - \widehat{D}(\rho \Vert \sigma)\\
        &\le p (\widehat{D}(\rho_1 \Vert \sigma_1) - \widehat{D}(\rho_1 \Vert \sigma)) \\ 
        & + (1 - p)(\widehat{D}(\rho_2 \Vert \sigma_2) - \widehat{D}(\rho_2 \Vert \sigma)) + \hat{c}_0 h(p) \, .
    \end{aligned}
\end{IEEEeqnarray}
Indeed, as of Lemmas \ref{lem:alternative_representation_bs_entropy} and then \ref{lem:almost_concavity_xlogx} with $A_1 = \sigma^{-1/2} \rho_1 \sigma^{-1/2}$, $A_2 = \sigma^{-1/2} \rho_2 \sigma^{-1/2}$ respectively, we can prove
\begin{IEEEeqnarray*}{ll}
    - & \widehat{D}(\rho \Vert \sigma) = \tr[\sigma \left(-\sigma^{-1/2} \rho \sigma^{-1/2} \log(\sigma^{-1/2} \rho \sigma^{-1/2}) \right)]\\
    &\le p \tr[\sigma \left(-\sigma^{-1/2} \rho_1 \sigma^{-1/2} \log(\sigma^{-1/2} \rho_1 \sigma^{-1/2}) \right)]\\
    &\hspace{0cm} + (1 - p) \tr[\sigma \left(-\sigma^{-1/2} \rho_2 \sigma^{-1/2} \log(\sigma^{-1/2} \rho_2 \sigma^{-1/2})\right)]\\ 
    &\hspace{0cm} + h_{A_1, A_2}(p)\\[2mm]
    &= -p \widehat{D}(\rho_1 \Vert \sigma) - (1 - p) \widehat{D}(\rho_2 \Vert \sigma) + h_{A_1, A_2}(p) \, .
\end{IEEEeqnarray*}
At last we can estimate $\tr[A_j] = \tr[\sigma^{-1}\rho_j] \le \norm{\sigma^{-1}}_\infty \le \hat{c}_0$ for $j = 1, 2$ using Hölder's inequality, giving us $h_{A_1, A_2}(p) \le \hat{c}_0 h(p)$.

We now have to estimate terms of the form $\widehat{D}(\rho_j \Vert \sigma_j) - \widehat{D}(\rho_j \Vert \sigma)$ for $j = 1, 2$. This is done using the Peierls-Bogolubov inequality \cite{AndoHiai_PeierlsBogolubov_1998} and the multivariate trace inequalities of Sutter et al. \cite{SutterBertaTomamichel-Multivariate-2017}:
\begin{IEEEeqnarray}{ll}
    & \widehat{D}(\rho_j \Vert \sigma_j)  - \widehat{D} (\rho_j \Vert \sigma) \nonumber\\ 
    &= \tr[\rho_j\left(\log(\rho_j^{1/2} \sigma_j^{-1} \rho_j^{1/2}) - \log(\rho_j^{1/2} \sigma^{-1} \rho_j^{1/2})\right)] \nonumber\\[1.3mm]
    &\le \tr[\exp\left(\log(\rho_j) + \log(\rho_j^{1/2}\sigma_j^{-1}\rho_j^{1/2}) - \log(\rho_j^{1/2}\sigma^{-1}\rho_j^{1/2})\right)]\nonumber\\[1.3mm]
    &\le \tr[\exp\left(\log(\rho_j) + \log(\rho_j^{1/2}\sigma_j^{-1}\rho_j^{1/2}) + \log(\rho_j^{-1/2}\sigma\rho_j^{-1/2})\right)]\nonumber\\
    & \le \log \alpha\left( \rho_{j}, \rho_{j}^{1/2}\sigma_{j}^{-1}\rho_{j}^{1/2}, \rho_{j}^{-1/2} \sigma \rho^{-1/2}\right)\nonumber\\
    &= \begin{cases}
        \log(p + (1 - p) \hat{c}_1) & j = 1\\
        \log((1 - p) + p \hat{c}_2) & j = 2
    \end{cases} \, . \label{eq:eq_int_ineq_bs_entropy}
\end{IEEEeqnarray}
In the third line, we use that 
\begin{equation*}
    -\log(\rho_j^{1/2} \sigma^{-1} \rho_j^{1/2}) \le \log(\rho_j^{-1/2} \sigma \rho_j^{-1/2})
\end{equation*}
which is true since for $P_{\rho}$ the projection on the support of $\rho$, we have 
\begin{equation*}
    P_{\rho} (P_{ \rho} \sigma P_{\rho})^{-1} P_{\rho} \le P_{\rho} \sigma^{-1} P_{\rho} \,,
\end{equation*}
as $x \to x^{-1}$ is operator convex and hence fulfills the Sherman-Davis inequality \cite[Theorem 4.19]{Carlen-TraceInequalities-2009}. Note that $\sigma$ is invertible and that by $(P_{ \rho} \sigma P_{\rho})^{-1}$ we mean the Moore-Penrose pseudoinverse. We find
\begin{IEEEeqnarray*}{ll}
    -\log(\rho_j^{1/2} \sigma^{-1} \rho_j^{1/2}) &= -\log(\rho_j^{1/2} P_{\rho} \sigma^{-1} P_{\rho} \rho_j^{1/2}) \\
    &\le -\log(\rho_j^{1/2} P_{\rho} (P_{\rho} \sigma P_{ \rho})^{-1} P_{\rho} \rho_j^{1/2})\\
    &= \log(\rho_j^{-1/2}P_{\rho} \sigma P_{\rho}\rho_j^{-1/2})\\
    &= \log(\rho_j^{-1/2} \sigma \rho_j^{-1/2}) \, .
\end{IEEEeqnarray*}
The argument why the inequalities in \eqref{eq:eq_int_ineq_bs_entropy} hold in the case of $\rho_j$ not being full rank is simpler than in the case of the corresponding inequality for the Umegaki relative entropy. For the BS-entropy, we can already restrict \eqref{eq:eq_int_ineq_bs_entropy} to the support of $\rho_j$ as all operators involved, $\rho_j$, $\rho_j^{1/2} \sigma_j^{-1} \rho_j^{1/2}$ and $\rho_j^{1/2} \sigma^{-1} \rho_j^{1/2}$, commute with the projection onto this support. In the last step we split $\sigma$ and evaluated the first term to $p$ in case $j = 1$ or the second term in case $j = 2$ to $(1 - p)$ and left the other one untouched, respectively. This concludes the proof. 

\hfill\IEEEQED

\subsection{Proof of Corollary~\ref{cor:uncertainty_relations}}

First, note that, given $\rho_{AM} \in \mathcal{S}(\mathcal{H}_{AM})$, for $\mathcal{Z} \in \{ \mathcal{X} , \mathcal{Y} \}$ the following holds:
\begin{IEEEeqnarray}{rCl}
    \begin{aligned}
        & H(Z | M)_{(E_{\mathcal{Z}}\otimes \operatorname{id}_M)(\rho_{AM})} \nonumber \\ 
        &= - D \left( (E_{\mathcal{Z}}\otimes \operatorname{id}_M)(\rho_{AM}) \Big\| \frac{\identity_A}{d_A} \otimes \rho_M \right)  + \log d_A \, ,
    \end{aligned}
\end{IEEEeqnarray}
and $(E_{\mathcal{M}}\otimes \operatorname{id}_M)(\rho_{AM}) = \frac{\identity_A}{d_A} \otimes \rho_M$. Moreover, by \cite[Lemma 3.4]{junge2019stability},
\begin{IEEEeqnarray}{rCl}
    \begin{aligned}
        & D \left( (E_{\mathcal{Z}}\otimes \operatorname{id}_M)(\rho_{AM}) \Big\| (E_{\mathcal{M}}\otimes \operatorname{id}_M)(\rho_{AM}) \right) \nonumber \\ 
        &\phantom{asasadsdasdasd}=  D \left( \rho_{AM} \Big\| (E_{\mathcal{M}}\otimes \operatorname{id}_M)(\rho_{AM}) \right) \\
        &\phantom{asasadsdasdsdasd} - D \left( \rho_{AM} \Big\| (E_{\mathcal{Z}}\otimes \operatorname{id}_M)(\rho_{AM}) \right) \, .     
    \end{aligned}
\end{IEEEeqnarray}
Therefore, we have
\begin{IEEEeqnarray}{rCl}
    \begin{aligned}
        & H(X | M)_{(E_{\mathcal{X}}\otimes \operatorname{id}_M)(\rho_{AM})} + H(Y | M)_{(E_{\mathcal{Y}}\otimes \operatorname{id}_M)(\rho_{AM})} \nonumber \\ 
        &= D \left( \rho_{AM} \Big\| (E_{\mathcal{X}}\otimes \operatorname{id}_M)(\rho_{AM}) \right)  \\ 
        & \; \; \; + D \left( \rho_{AM} \Big\| (E_{\mathcal{Y}}\otimes \operatorname{id}_M)(\rho_{AM}) \right)   \\
        & \; \; \; - 2  D \left( \rho_{AM} \Big\| (E_{\mathcal{M}}\otimes \operatorname{id}_M)(\rho_{AM}) \right) + 2 \log d_A \, .
    \end{aligned}
\end{IEEEeqnarray}
Additionally, it is not difficult to see that 
\begin{IEEEeqnarray}{rCl}
    \begin{aligned}
        &  D \left( \rho_{AM} \Big\| ((E_{\mathcal{X}} \circ E_{\mathcal{Y}} )\otimes \operatorname{id}_M)(\rho_{AM}) \right) \\
        & = D \left( \rho_{AM} \Big\| (E_{\mathcal{X}}\otimes \operatorname{id}_M)(\rho_{AM})  \right) \nonumber \\ 
        & \; \; \; + D \left( (E_{\mathcal{X}}\otimes \operatorname{id}_M)(\rho_{AM}) \Big\| ((E_{\mathcal{X}} \circ E_{\mathcal{Y}} )\otimes \operatorname{id}_M)(\rho_{AM}) \right)   \\
        &  \leq D \left( \rho_{AM} \Big\| (E_{\mathcal{X}}\otimes \operatorname{id}_M)(\rho_{AM})  \right) \\
        & \; \; \; + D \left( \rho_{AM} \Big\| (E_{\mathcal{Y}}\otimes \operatorname{id}_M)(\rho_{AM})  \right)  \, ,
    \end{aligned}
\end{IEEEeqnarray}
also as a consequence of \cite[Lemma 3.4]{junge2019stability} for the first equality and the data-processing inequality for the relative entropy.  Furthermore, by Eq. \eqref{eq:cont_bound_rel_ent_second_input}, we have
\begin{IEEEeqnarray}{rCl}
    \begin{aligned}
        &\left| D \left( \rho_{AM} \Big\| \frac{\identity_A}{d_A} \otimes \rho_M \right)  - D \left( \rho_{AM} \Big\| ((E_{\mathcal{X}} \circ E_{\mathcal{Y}} )\otimes \operatorname{id}_M)(\rho_{AM}) \right)   \right| \nonumber \\ 
        & \;  \leq \frac{-2 \log m + \frac{1}{m}}{1- m}\norm{\frac{\identity_A}{d_A} \otimes \rho_M  - ((E_{\mathcal{X}} \circ E_{\mathcal{Y}} )\otimes \operatorname{id}_M)(\rho_{AM}) }_{1}^{1/2} \, ,
    \end{aligned}
\end{IEEEeqnarray}
Note that we can rewrite the last term on the right-hand side as: 
\begin{IEEEeqnarray}{rCl}
    \begin{aligned}
        &\norm{\frac{\identity_A}{d_A} \otimes \rho_M  - ((E_{\mathcal{X}} \circ E_{\mathcal{Y}} )\otimes \operatorname{id}_M)(\rho_{AM}) }_{1}  \nonumber \\ 
        &\phantom{asasdasdadasda} = \norm{  ((E_{\mathcal{M}}- E_{\mathcal{X}} \circ E_{\mathcal{Y}} )\otimes \operatorname{id}_M)(\rho_{AM}) }_1 \, ,
    \end{aligned}
\end{IEEEeqnarray}
A similar term was computed e.g. in \cite[Section 4.1]{bardet2020approximate}. Following analogous ideas, we have:
\begin{IEEEeqnarray}{rCl}
    \begin{aligned}
        &  ((E_{\mathcal{X}} \circ E_{\mathcal{Y}} ) \otimes \text{id}_M) (\rho_{AM}) \\
        &= \underset{x,y}{\sum} |e_x^{(\mathcal{X})} \rangle \langle e_x^{(\mathcal{X})}| e_y^{(\mathcal{Y})}  \rangle \langle e_y^{(\mathcal{Y})} | \rho_{AM} | e_y^{(\mathcal{Y})} \rangle \langle e_y^{(\mathcal{Y})}  | e_x^{(\mathcal{X})} \rangle \langle e_x^{(\mathcal{X})} |   \nonumber \\ 
        &  = \underset{x,y}{\sum}  | \langle e_x^{(\mathcal{X})} | e_y^{(\mathcal{Y})}  \rangle |^2   | e_x^{(\mathcal{X})} \rangle \langle e_x^{(\mathcal{X})} |   \otimes  \langle e_y^{(\mathcal{Y})} | \rho_{AM} | e_y^{(\mathcal{Y})} \rangle  \, ,
    \end{aligned}
\end{IEEEeqnarray}
and
\begin{IEEEeqnarray}{rCl}
    \begin{aligned}
        &  (E_{\mathcal{M}} \otimes  \text{id}_M) (\rho_{AM})  = \frac{1}{d_A}\underset{x,y}{\sum}   | e_x^{(\mathcal{X})} \rangle \langle e_x^{(\mathcal{X})} |   \otimes  \langle e_y^{(\mathcal{Y})} | \rho_{AM} | e_y^{(\mathcal{Y})} \rangle  \nonumber  \, ,
    \end{aligned}
\end{IEEEeqnarray}
Then,
\begin{IEEEeqnarray}{rCl}
    \begin{aligned}
        &\norm{  ((E_{\mathcal{M}}- E_{\mathcal{X}} \circ E_{\mathcal{Y}} )\otimes \operatorname{id}_M)(\rho_{AM}) }_1 \nonumber \\ 
        & =  \Bigg\|\underset{x,y}{\sum}  \left(\frac{1}{d_A} -  | \langle e_x^{(\mathcal{X})} | e_y^{(\mathcal{Y})}  \rangle |^2   \right) | e_x^{(\mathcal{X})} \rangle \langle e_x^{(\mathcal{X})} |   \\&\quad\otimes  \langle e_y^{(\mathcal{Y})} | \rho_{AM} | e_y^{(\mathcal{Y})} \rangle  \Bigg\|_1 \\
        & \leq \underset{x,y}{\sum}  \left| \frac{1}{d_A} -  | \langle e_x^{(\mathcal{X})} | e_y^{(\mathcal{Y})}  \rangle |^2   \right| \norm{\langle e_y^{(\mathcal{Y})} | \rho_{AM} | e_y^{(\mathcal{Y})} \rangle }_1 \\
        & \leq \underset{x}{\sum} \, \underset{y}{\text{max}} \left| \frac{1}{d_A} -  | \langle e_x^{(\mathcal{X})} | e_y^{(\mathcal{Y})}  \rangle |^2   \right|  \, ,
    \end{aligned}
\end{IEEEeqnarray}
Additionally, note that for $\Phi \in \mathcal{B}(\mathcal{H}_A)$,
\begin{IEEEeqnarray}{rCl}
    \begin{aligned}
        &  (E_{\mathcal{X}} \circ E_{\mathcal{Y}} ) (\Phi) \\
        &= \underset{x,y}{\sum} |e_x^{(\mathcal{X})} \rangle \langle e_x^{(\mathcal{X})}| e_y^{(\mathcal{Y})}  \rangle \langle e_y^{(\mathcal{Y})} | \Phi | e_y^{(\mathcal{Y})} \rangle \langle e_y^{(\mathcal{Y})}  | e_x^{(\mathcal{X})} \rangle \langle e_x^{(\mathcal{X})} |   \nonumber \\ 
        &  \geq \underset{x,y}{\text{min}}  | \langle e_x^{(\mathcal{X})} | e_y^{(\mathcal{Y})}  \rangle |^2  \; \text{tr}[\Phi] \identity_A \, .
    \end{aligned}
\end{IEEEeqnarray}
Moreover,
\begin{equation*}
    \rho_{AM} \leq e^{D_{\text{max}}\big(\rho_{AM} \| \frac{\identity_A}{d_A} \otimes \rho_M \big)} \frac{\identity_A}{d_A} \otimes \rho_M  \leq e^{2 \log d_A}  \frac{\identity_A}{d_A} \otimes \rho_M  \, .
\end{equation*}
Therefore, since $m$ should satisfy  $ m \rho_{AM} \leq ( (E_{\mathcal{X}} \circ  E_{\mathcal{Y}} ) \otimes \text{id}_M)(\rho_{AM})$ and $ m \rho_{AM} \leq \frac{\identity_A}{d_A} \otimes \rho_M $, then it can be taken to be 
\begin{equation}
    m:= \text{min} \left\{ \frac{1}{(d_A)^2}  , \frac{1}{d_A}\underset{x,y}{\text{min}} | \langle e_x^{(\mathcal{X})} | e_y^{(\mathcal{Y})}  \rangle |^2 \right\} \, .
\end{equation}
This concludes the proof.

\hfill\IEEEQED

\end{document}